\documentstyle[12pt,emulateapj]{article}

\input epsf.sty

\def\lsim{\mathrel{\rlap{\lower 4pt \hbox{\hskip 1pt $\sim$}}\raise 1pt \hbox
        {$<$}}}
\def\gsim{\mathrel{\rlap{\lower 4pt \hbox{\hskip 1pt $\sim$}}\raise 1pt \hbox
        {$>$}}}

\newcommand{\etal}{et~al.\ }
\newcommand{\eg}{e.g.\ }

\newcommand{\Msun}{M$_{\odot}$}

\newcommand{\kms}{km~s$^{-1}$}
\newcommand{\Mej}{$M_{\rm ej}$}
\newcommand{\HeI}{He~{\sc i}}
\newcommand{\CI}{C~{\sc i}}

\newcommand{\OI}{O~{\sc i}}

\newcommand{\NaI}{Na~{\sc i}}
\newcommand{\MgII}{Mg~{\sc ii}}

\newcommand{\SiII}{Si~{\sc ii}}

\newcommand{\CaII}{Ca~{\sc ii}}

\newcommand{\FeII}{Fe~{\sc ii}}

\newcommand{\Fefs}{$^{56}$Fe}
\newcommand{\Cofs}{$^{56}$Co}
\newcommand{\Nifs}{$^{56}$Ni}

\begin{document}

\title{The Type Ic Hypernova SN 2002ap}

\author{P. A. Mazzali\altaffilmark{1,2,3}, J. Deng\altaffilmark{1,2},
K. Maeda\altaffilmark{2}, K. Nomoto\altaffilmark{1,2}, 
H. Umeda\altaffilmark{1,2}, K. Hatano\altaffilmark{1,2},
K.~Iwamoto\altaffilmark{4},
Y. Yoshii\altaffilmark{1,5}, Y. Kobayashi\altaffilmark{6},
T. Minezaki\altaffilmark{5}, M. Doi\altaffilmark{5}, 
K. Enya\altaffilmark{5},  H. Tomita\altaffilmark{2,5},
S.J.~Smartt\altaffilmark{7}, 
K. Kinugasa\altaffilmark{8}, H. Kawakita\altaffilmark{8}, 
K. Ayani\altaffilmark{9}, T. Kawabata\altaffilmark{9}, 
H. Yamaoka\altaffilmark{10},
Y.L.~Qiu\altaffilmark{11},
K. Motohara\altaffilmark{5}, 
C. L. Gerardy\altaffilmark{12}, R. Fesen\altaffilmark{12},
K. S. Kawabata\altaffilmark{6}, M. Iye\altaffilmark{6,13}, 
N.~Kashikawa\altaffilmark{6}, G. Kosugi\altaffilmark{14},
Y. Ohyama\altaffilmark{14}, M. Takada-Hidai\altaffilmark{15}, 
G. Zhao\altaffilmark{11}, 
R.~Chornock\altaffilmark{16}, A. V. Filippenko\altaffilmark{16},
S. Benetti\altaffilmark{17}, and M. Turatto\altaffilmark{17} }

\altaffiltext{1}{Research Center for the Early Universe, Univ. of Tokyo,
	Bunkyo-ku, Tokyo, Japan}
\altaffiltext{2}{Dept. of Astronomy, Univ. of Tokyo, Bunkyo-ku, Tokyo, Japan }
\altaffiltext{3}{Osservatorio Astronomico, Via Tiepolo, 11, 34131 Trieste, 		
	Italy}
\altaffiltext{4}{Dept. of Physics, Nihon Univ., Chiyoda-ku, Tokyo, Japan}
\altaffiltext{5}{Inst. of Astronomy, Univ. of Tokyo, Mitaka, Tokyo, Japan }
\altaffiltext{6}{National Astronomical Observatory, Mitaka, Tokyo, Japan}
\altaffiltext{7}{Institute of Astronomy, University of Cambridge, Madingley Rd., 	
	Cambridge, England}
\altaffiltext{8}{Gunma Astronomical Observatory, Takayama, Gunma, Japan}
\altaffiltext{9}{Bisei Astronomical Observatory, Bisei, Okayama, Japan}
\altaffiltext{10}{Dept. of Physics, Kyushu Univ., Chuo-ku, Fukuoka, Japan}
\altaffiltext{11}{National Astronomical Observatories, 
	Chinese Academy of Sciences, Beijing, China }
\altaffiltext{12}{Dept. of Physics \& Astronomy, Dartmouth College, Hanover, NH,
	 USA }
\altaffiltext{13}{Dept. of Astronomy, Graduate University for Advanced Studies,
        Mitaka, Tokyo, Japan}
\altaffiltext{14}{Subaru Telescope, National Astronomical Observatory of Japan,
    Hilo, HW, USA}
\altaffiltext{15}{Liberal Arts Education Center, Tokai University, Hiratsuka, 	
	Kanagawa, Japan}
\altaffiltext{16}{Department of Astronomy, University of California, Berkeley,
   CA, USA}
\altaffiltext{17}{Osservatorio Astronomico, Vicolo dell'Osservatorio, 2, Padova,
		Italy}

\begin{abstract}

Photometric and spectroscopic data of the energetic Type Ic supernova (SN)
2002ap are presented, and the properties of the SN are investigated through
models of its spectral evolution and its light curve.  The SN is
spectroscopically similar to the ``hypernova" SN~1997ef.  However, its kinetic
energy [$\sim (4-10) \times 10^{51}$ erg] and the mass ejected (2.5--5 \Msun)
are smaller, resulting in a faster-evolving light curve.  The SN synthesized
$\sim 0.07$\Msun\ of \Nifs, and its peak luminosity was similar to that of
normal SNe.  Brightness alone should not be used to define a hypernova, whose
defining character, namely very broad spectral features, is the result of a
high kinetic energy.  The likely main-sequence mass of the progenitor star was
20--25 \Msun, which is also lower than that of both hypernovae SNe 1997ef and
1998bw.  SN~2002ap appears to lie at the low-energy and low-mass end of the
hypernova sequence as it is known so far. Observations of the nebular spectrum,
which is expected to dominate by summer 2002, are necessary to confirm these
values.

\end{abstract}


\newpage

\section{Introduction}

One of the most interesting recent developments in the study of supernovae (SNe)
is the discovery of some very energetic Type Ic SNe (SNe~Ic; see Filippenko 1997
for a general review), whose kinetic energy ($KE$) exceeds $10^{52}$\,erg, about
10 times the $KE$ of normal core-collapse SNe (hereafter $E_{51} =
10^{51}$\,erg).  The most luminous and powerful of these objects, SN~1998bw, was
probably linked to GRB 980425 (Galama \etal 1998), thus establishing for the
first time a connection between the enigmatic gamma-ray bursts (GRBs) and the
well-studied phenomenon of core-collapse SNe.  However, SN~1998bw was
exceptional for a SN~Ic: it was as luminous at peak as a SN~Ia, indicating that
it synthesized $\sim 0.5$ \Msun\ of \Nifs, and its $KE$ was estimated at $\sim 3
\times 10^{52}$ erg (Iwamoto \etal 1998; Woosley \etal 1999).  Because of its
large $KE$, SN~1998bw was classified as a ``hypernova".  Signatures of asymmetry
were detected in SN~1998bw, such as polarization (Patat \etal 2001) and peculiar
nebular-line profiles (Mazzali \etal 2001; Maeda \etal 2002).

Subsequently, other ``hypernovae" of Type Ic have been discovered or
recognised, such as SN~1997ef (Iwamoto \etal 2000; Mazzali \etal 2000) and
SN~1997dq (Matheson \etal 2001; Mazzali \etal 2002, in preparation), although
their $KE$ was not as large as that of SN~1998bw, and they did not appear to be
associated with GRBs.  The analysis of these various objects suggests that the
$KE$ may be related to the progenitor's main-sequence mass, which was $\sim 40$
\Msun\ for SN~1998bw and $\sim 30$ \Msun\ for SN~1997ef.  These values place
hypernovae at the high-mass end of SN progenitors.  Another possible hypernova,
although of Type IIn, was SN~1997cy, which was also estimated to have a large
mass ($\sim 25$ \Msun; Germany \etal 2000; Turatto \etal 2000). It is not yet
clear whether the large mass is the discriminating factor for the birth of a
hypernova or the connection with a GRB.

In this paper we present a first analysis of the properties of the recently
discovered SN~Ic 2002ap in M74 (Hirose 2002).  The SN was immediately recognised
as a hypernova from its broad spectral features (Kinugasa \etal 2002; Meikle
\etal 2002; Gal-Yam \etal 2002a; Filippenko \& Chornock 2002).  This indicates
high velocities in the ejected material, which is the typical signature of
hypernovae.  It was therefore followed from several observatories, and the
relative proximity also favored observations with small telescopes.  Luckily,
the SN was discovered very soon after it exploded:  the discovery date was
January 29, while the SN was not detected on January 25 (Nakano \etal 2002).
This is among the earliest any SN has been observed, with the obvious
exceptions of SN~1987A and SN~1993J.

Figure 1 shows the near-maximum spectra of SN~2002ap, of the hypernovae SNe
1998bw and 1997ef, and of the normal SN~Ic 1994I.  If line width is the
distinguishing feature of a hypernova, then clearly SN~2002ap is a hypernova, as
its spectrum resembles that of SN~1997ef much more than that of SN~1994I.  Line
blending in SN~2002ap and SN~1997ef is comparable.  However, some individual
features that are clearly visible in SN~1994I but completely blended in
SN~1997ef can at least be discerned in SN~2002ap (\eg the \NaI--\SiII\ blend
near 6000~\AA\ and the \FeII\ lines near 5000~\AA).  Therefore,
spectroscopically SN~2002ap appears to be located just below SN~1997ef in a
``velocity scale," but far above SN~1994I.

This appears to be confirmed by the light curve.  Figure 2 shows the $V$-band
light curves of the same four SNe as in Figure 1. SN~2002ap reached $V$ maximum
on about February 8 at $V = 12.3$ mag.  SN~2002ap peaks earlier than both hypernovae 
1998bw and 1997ef, but later than the normal SN~1994I, suggesting an intermediate value of the
ejecta mass \Mej.

Using a distance to M74 of 8~Mpc ($\mu
= 29.5$ mag; Sharina \etal 1996), and a combined Galaxy and M74 reddening of
$E(B-V) = 0.09$ mag (estimated from a Subaru HDS spectrum; Takada-Hidai et al.
2002, in preparation), the absolute magnitude is $M_V = -17.4$.  This is
comparable to SN~1997ef and fainter than SN~1998bw by almost 2 mag.  Since peak
brightness depends on the ejected \Nifs\ mass, SNe~2002ap, 1997ef, and 1994I
appear to have synthesized similar amounts of it.  Estimates were $\sim 0.07$
\Msun\ for SN~1994I (Nomoto \etal 1994) and 0.13 \Msun\ for SN~1997ef (Mazzali
\etal 2000).  

Section 2 of this paper describes how an explosion model was selected that gives
reasonable fits to the spectra of SN~2002ap, thus establishing the properties of
the SN.  In \S~3 we show synthetic light curves obtained with our best models,
which confirm the results of the spectroscopic analysis.  In \S~4 we discuss
implications for the progenitor of SN~2002ap.

\section{Spectroscopic Models} 

Iwamoto \etal (1998) showed that synthetic light curves of SNe~Ic are degenerate
with respect to particular sets of the parameters \Mej, $KE$, and the opacity.
Spectra must also be used to determine those parameters, as their dependence is
different.  Therefore we started the analysis from the spectra.  We used the
Monte-Carlo SN spectrum synthesis code described by Mazzali \& Lucy (1993) and
modified as in Lucy (1999) and Mazzali (2000).

We tested several versions of the hypernova explosion models CO138 (used for
SN~1998bw; Iwamoto \etal 1998) and CO100 (used for SN~1997ef; Mazzali \etal
2000).  Both of the models turned out to be too massive.  In fact, since
SN~2002ap rises very rapidly in brightness, the epochs of its spectra are
smaller than those of similar-looking spectra of both SNe 1997ef and 1998bw.  
In particular, the earliest spectrum of SN~2002ap, taken on January 30, has an
epoch of at most 4 days, but it resembles the spectra of SNe 1997ef and 1998bw
at epochs of about 8 days, indicating similar temperatures and expansion
velocities.  The densities in the massive models at such an early epoch are much
too high, and no good solution can be obtained unless most of the mass is
composed of He, which contributes very little both to the electron density and
to line opacity.

As the subsequent evolution of SN~2002ap more closely resembles that of
SN~1997ef (albeit at a faster rate) than that of SN~1998bw, we selected model
CO100 and rescaled it to lower masses, and hence lower $KE$.  We tried a model
with half the mass (model CO100/2:  \Mej = 4.8 \Msun; $E_{51} = 8$) and one with
a quarter of the mass (model CO100/4:  \Mej = 2.4 \Msun; $E_{51} = 4$).  Model
CO100/2 is too massive, while model CO100/4 appears to be roughly appropriate
for SN~2002ap.  Figure 3 shows a series of fits obtained with that model.

Also, the earliest spectrum (Meikle \etal 2002) requires a large degree of line
blending, as shown by the apparent lack of any separation between \OI\
$\lambda$7774 and the \CaII\ near-infrared triplet.  This requires the presence
of sufficient material at $v > 30,000$ \kms.  In the case of SN~1997ef, this
blending was obtained by adding an outer ``flat" part to the density profile
(Mazzali \etal 2000).  Blending in the \OI--\CaII\ feature in the earliest
observed spectrum of SN~1997ef line was weaker, but the epoch of this spectrum
is later than the first spectrum of SN~2002ap.  We also modified model CO100
slightly by limiting the velocities to 65,000 \kms, to match the observations.

Because the first spectrum was obtained so early, it was possible to establish
its age fairly accurately using synthetic spectra.  Age and velocity must in
fact combine to give an acceptable temperature, but velocity can be checked
against the blueshift of the line absorptions.  For January 30 we find $t = 2.0
\pm 0.5$ days.

The first spectrum has a photospheric velocity $v_{\rm ph} = 30,000$ \kms, and
thus the photosphere is located just below the outer flat extension of the
density profile.  Hence line blending is very strong, and the spectrum
resembles that of SN~1998bw.  But already the next spectrum, with $t = 3.5$
days, has $v_{\rm ph} = 20,500$ \kms, which is in the steep part of the density
distribution, so blending becomes less severe.  Note that the \CaII\ near-IR
triplet can blend with the \OI\ line even with a very small abundance of Ca at
high velocity, because it is a very easy line to excite.  We see that $v_{\rm
ph}$ continues to decrease with time, but the rate of decrease becomes smaller
as the light curve reaches peak, which is expected as the photosphere then
enters the inner flat part of the density distribution.

Oxygen dominates the composition in all the spectra we have modeled.  However,
the earliest spectrum contains $\sim 30$\% He by mass, suggesting that the SN
exploded with about 0.05 \Msun\ left from the He envelope.  This appears to be
consistent with the early detection of an absorption at about 1.03 $\mu$m,
which could be \HeI\ $\lambda$10830, although \MgII\ and \CI\ could also
contribute to it (Fig. 3; Motohara \etal 2002).  Synthetic spectra can only
reproduce this feature if significant nonthermal ionization of He is active,
which is also not unexpected. The abundances of Si and other intermediate-mass
elements, and those of the Fe-group elements, increase with depth in the
ejecta, which is in accord with expectations.

Several problems remain in the models.  The synthetic \OI\ and \CaII\
absorptions are too blue at the more advanced epochs, suggesting that the
density profile we have adopted may overestimate the mass at high velocities.
Also, the ratio of the \FeII\ lines in the absorption near 4800~\AA\ is not
correct.  Further, more refined analysis is therefore needed.  However, we do
not expect that our general results will change significantly.

\section{Light-Curve Models} 

We constructed the UVOIR bolometric light curve of SN~2002ap mainly from the
$UBV(RI)_cJHK$ photometry obtained with the MAGNUM telescope (Yoshii \etal
2002). We also included 5 data points based on observations from Wise
Observatory (Gal-Yam \etal 2002b), the U.K. Infrared Telescope (Mattila \etal
2002), MDM (Gerardy \etal 2002, in preparation) and the Subaru Telescope
(Motohara et al. 2002).  We estimate the errors at 0.3 mag, mostly owing to the
uncertainty in the distance modulus ($\sim 0.2$ mag) and the nonstellar nature
of the spectrum.  The UVOIR magnitudes and the bolometric magnitudes derived
from the synthetic spectra agree to within 0.1 mag.  The IR flux accounts for a
significant fraction of the total flux, ranging from $\sim 20$\% on February 4
to $\sim 40$\% on March 10, while the UV flux observed by XMM-Newton on
February 3 (Pascual \etal 2002) contributes only $\sim 4$\%.

Synthetic light curves were computed with an LTE radiation hydrodynamics code
and a gray $\gamma$-ray transfer code (Iwamoto \etal 2000).  TOPS opacities
(Magee \etal 1995) were fitted to the ejecta models in order to find an
empirical relationship between the Rosseland mean and the electron-scattering
opacity.

Figure 4 shows the bolometric data and two synthetic light curves computed with
the same density structures used for spectrum synthesis.  The two models produce
very similar light curves, demonstrating the parameter degeneracy (\S 2).  Both
model light curves give a satisfactory fit to the observations, confirming the
results of the spectral analysis.

To reproduce the observed peak luminosity, a total of 0.07 \Msun of $^{56}$Ni
is used.  In order to achieve a rapid rise of the light curve, and in
particular to reproduce the earliest bolometric point on day 2, we had to mix
$^{56}$Ni out to high velocities, which introduces a local concentration of
$^{56}$Ni very close to the surface, as in the best-fit model for SN~1998bw
(Nakamura \etal 2001). Such significant outward mixing of the newly synthesized
$^{56}$Ni is consistent with our spectral analysis. 

Considering the uncertainties in the UVOIR bolometric light curve, we suggest
that $M(^{56}$Ni) = $0.07 \pm 0.02$ \Msun.  One should not be misled by the fact
that SN 2002ap is more luminous in the $V$ band than SN 1994I to conclude that
the former has a larger $^{56}$Ni mass.  In fact, the bolometric corrections
(B.C. $\equiv M_{\rm bol}-M_{\rm V}$) in SN~2002ap are quite large, ranging
from 0.5 to 0.3, during the first 20 days.  In contrast, SN 1994I shows a small
B.C. of 0.05 at the luminosity maximum (Iwamoto \etal 1994).  Large values for
the B.C. are also found in SN 1997ef (Mazzali \etal 2000).

\section{Discussion and Conclusions}

The spectral evolution of SN~2002ap appears to follow closely that of SN~1997ef,
at about twice the rate.  The spectra and the light curve of SN 2002ap can be
well reproduced by a model with ejected heavy-element mass \Mej = 2.5--5 \Msun\
and $E_{51}=4$--10.  Both \Mej\ and $KE$ are much smaller than those of SNe
1998bw and 1997ef (but they could be larger if a significant amount of He is
present).  The $^{56}$Ni mass is estimated to be $\sim 0.07$ \Msun, which is
similar to that of normal core-collapse SNe such as SNe 1987A and 1994I.

Although SN~2002ap appears to lie between classical core-collapse SNe and
hypernovae, it should be regarded as a hypernova because its kinetic energy is
distinctly higher than for classical core-collapse SNe.  In other words, the
broad spectral features that characterize hypernovae are the results of a high
kinetic energy.  Also, SN~2002ap was not more luminous than normal
core-collapse SNe.  Therefore brightness alone should not be used to
discriminate hypernovae from normal SNe, while the criterion should be a high 
kinetic energy. Further examples of hypernovae are necessary in order to
establish whether a firm boundary between the two groups exists.

For these values of $KE$, \Mej, and $M(^{56}$Ni), we can constrain the
progenitor's main-sequence mass $M_{\rm ms}$ and the remnant mass $M_{\rm rem}$.
Modeling the explosions of C+O stars with various masses, we obtain $M(^{56}$Ni)
as a function of the parameter set ($KE$, $M_{\rm CO}$, $M_{\rm rem} = $\Mej$ -
M_{\rm CO}$).  The model which is most consistent with our estimates of (\Mej,
$KE$) is one with $M_{\rm CO} \approx 5$\,\Msun, $M_{\rm rem} \approx
2.5$\,\Msun, and $E_{51} = 4$.  The 5.0 $M_\odot$ C+O core forms in a He core
of mass $M_\alpha = 7.0$ \Msun, corresponding to a main-sequence mass $M_{\rm
ms} \simeq 20$--25 \Msun.  The $M_{\rm ms} - M_\alpha$ relation depends on
convection and metallicity (e.g., Nomoto \& Hashimoto 1988; Umeda \& Nomoto
2002).  Given the non-detection of the progenitor in pre-discovery images of M74
(Smartt \etal 2002), the above mass range seems to be most consistent with the
progenitor being a member of an interacting binary system.

In order to determine the mass of O and that of \Nifs, whose decay to \Fefs\ via
\Cofs\ will be powering the nebular spectrum, it is very important that spectra
are taken in the fully nebular epoch.  There is already some evidence for net 
emission in the \CaII\ near-IR triplet in a February 21 spectrum, which has an
epoch of 24 days and looks similar to SN~1997ef on about day 42.  Eventually,
though, \OI\ $\lambda$6300 should become the strongest line.  Because of the
fast evolution of SN~2002ap, the time when the SN is again visible (late June
2002), corresponding to an epoch of 5 months, should already be suitable.

The estimated progenitor mass and explosion energy are both smaller than those
of previous ``hypernovae" such as SNe 1998bw and 1997ef, but larger than those
of normal core-collapse SNe such as SN 1999em.  There appears to be a
correlation between initial stellar mass and explosion energy, but further
observational examples are needed to establish the relation.  In particular, it
is unclear what fraction of massive stars with $M \gsim 20$ \Msun\ explode
energetically.  Massive core-collapse SNe with either a normal explosion energy
(e.g., SN~1984L; Swartz \& Wheeler 1991) or a very small one (SN~1997D; Turatto
\etal 1999) also appear to exist.

Given the estimated mass of the progenitor, binary interaction including the
spiral-in of a companion star (Nomoto \etal 2001) is probably required in order
for it to lose its hydrogen and some (or most) of its helium envelope.  This
would suggest that the progenitor was in a state of high rotation.  It is
possible that a high rotation rate and/or envelope ejection are also necessary
conditions for the birth of a hypernova.

The high explosion energy of SNe 1998bw and 1997ef may be related to the
formation of a black hole  through the extraction of energy from the black
hole-accretion disk system.  Also in the case of SN~2002ap, the expected remnant
mass ($\sim 2.5$ \Msun) exceeds the maximum neutron star mass, even though the
estimated main-sequence mass (20-25 \Msun) is not extremely large.

SN~2002ap was not apparently associated with a GRB.  This may actually be not so
surprising, since the explosion energy of SN~2002ap is about a factor of 5-10
smaller than that of SN~1998bw, as also indicated by the weak radio signature
(Berger \etal 2002).  The present data show no clear signature of asymmetry,
except perhaps for some polarization (Wang \etal 2002; Kawabata \etal 2002),
which is smaller than that of SN~1998bw.  This suggests that the degree of
asphericity is smaller in SN~2002ap and that the possible ``jet" may have been
weaker, which makes GRB generation more difficult.  It will therefore be
interesting to see whether the photosphere persists to very low velocities, as
it did in SN~1997ef, in apparent contradiction with one-dimensional explosion
models.

{\bf Acknowledgements.} We thank Peter Meikle for allowing us to use the 30 
January spectrum prior to publication. This work would not have been possible 
without the effort and enthusiasm of the staff at all the different 
observatories. This work has been supported in part by the grant-in-Aid for
Scientific Research (12640233, 14047206, 14540223) and COE research (07CE2002)
of the Ministry of Education, Science, Culture, Sports, and Technology in Japan.
J.S.~Deng is supported by a JSPS Postdoctoral Fellowship for Foreign 
Researchers.


\newpage
\begin{figure*}
\plotone{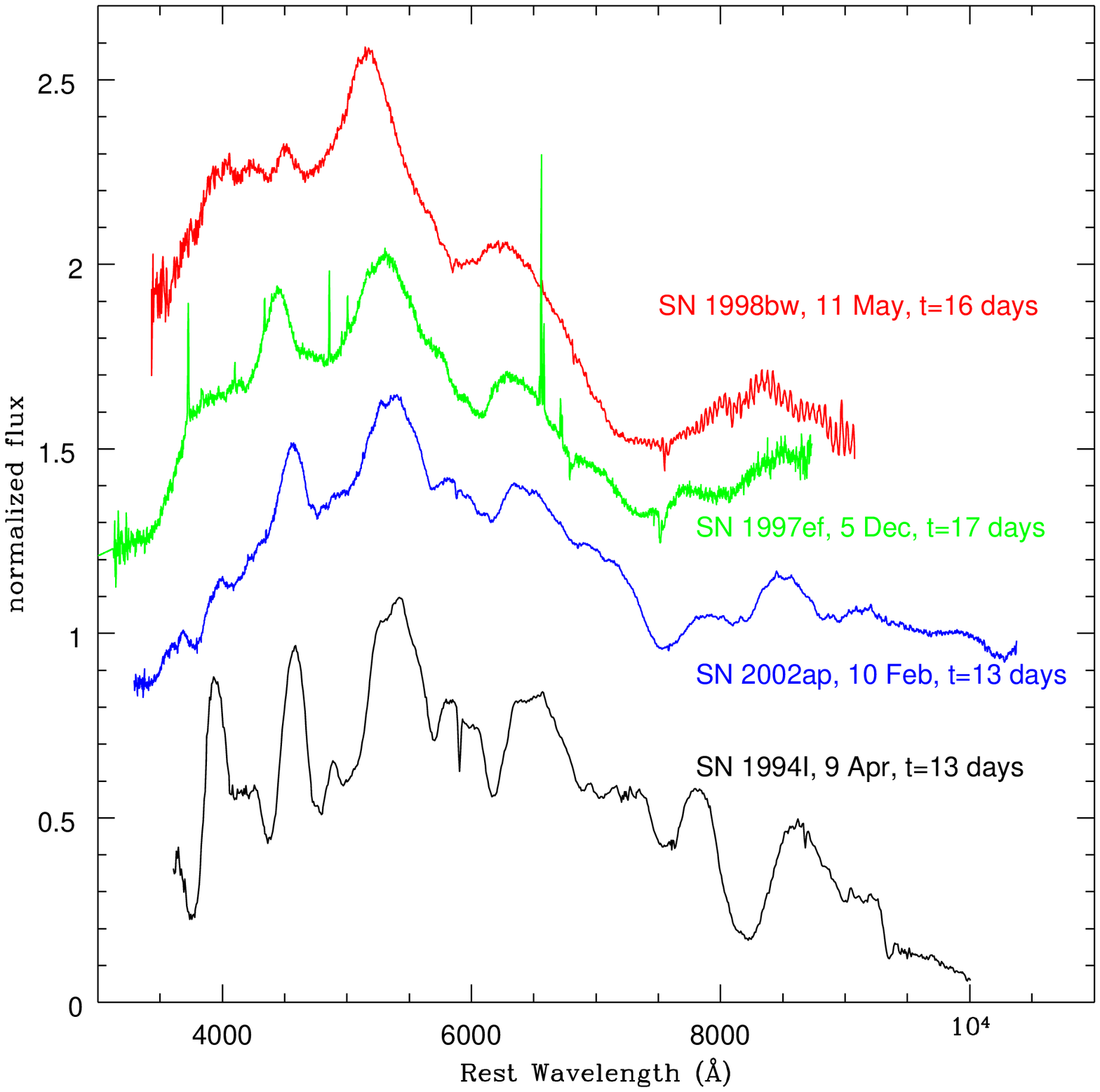}
\figcaption[.ps]{The near-maximum spectra of Type Ic SNe and hypernovae. }
\end{figure*}

\newpage
\begin{figure*}
\plotone{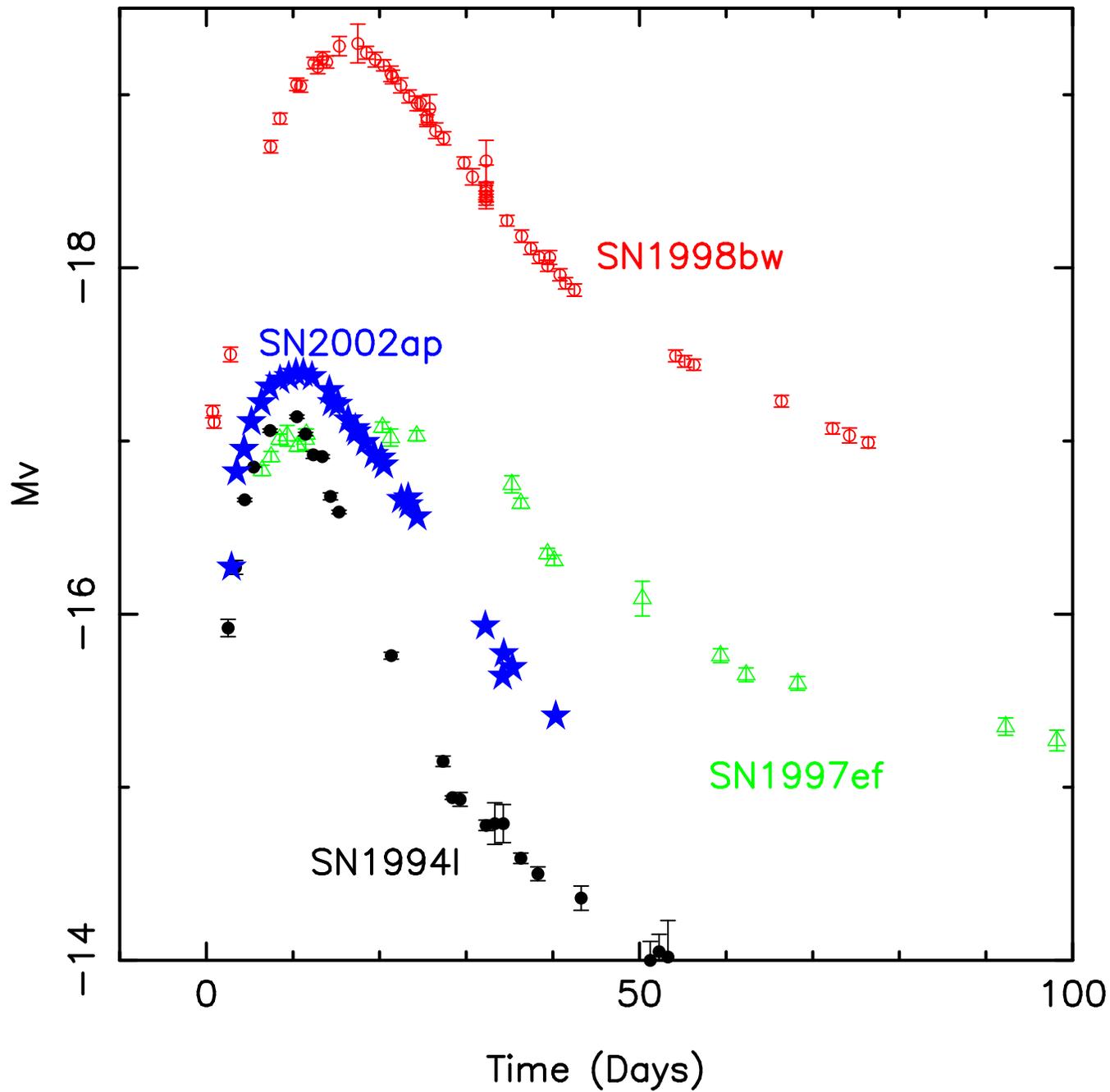}
\figcaption[.ps]{The observed $V$-band light curves of SNe 2002ap ({\em stars}),
1998bw ({\em open circles}), 1997ef ({\em open triangles}), 
and 1994I ({\em filled circles}). 
Most of the SN 2002ap data are taken from the MAGNUM
telescope (Yoshii \etal 2002) and the rest from the VSNET
(http://www.kusastro.kyoto-u.ac.jp/vsnet/). }
\end{figure*}

\newpage
\begin{figure*}
\epsscale{0.7}
\plotone{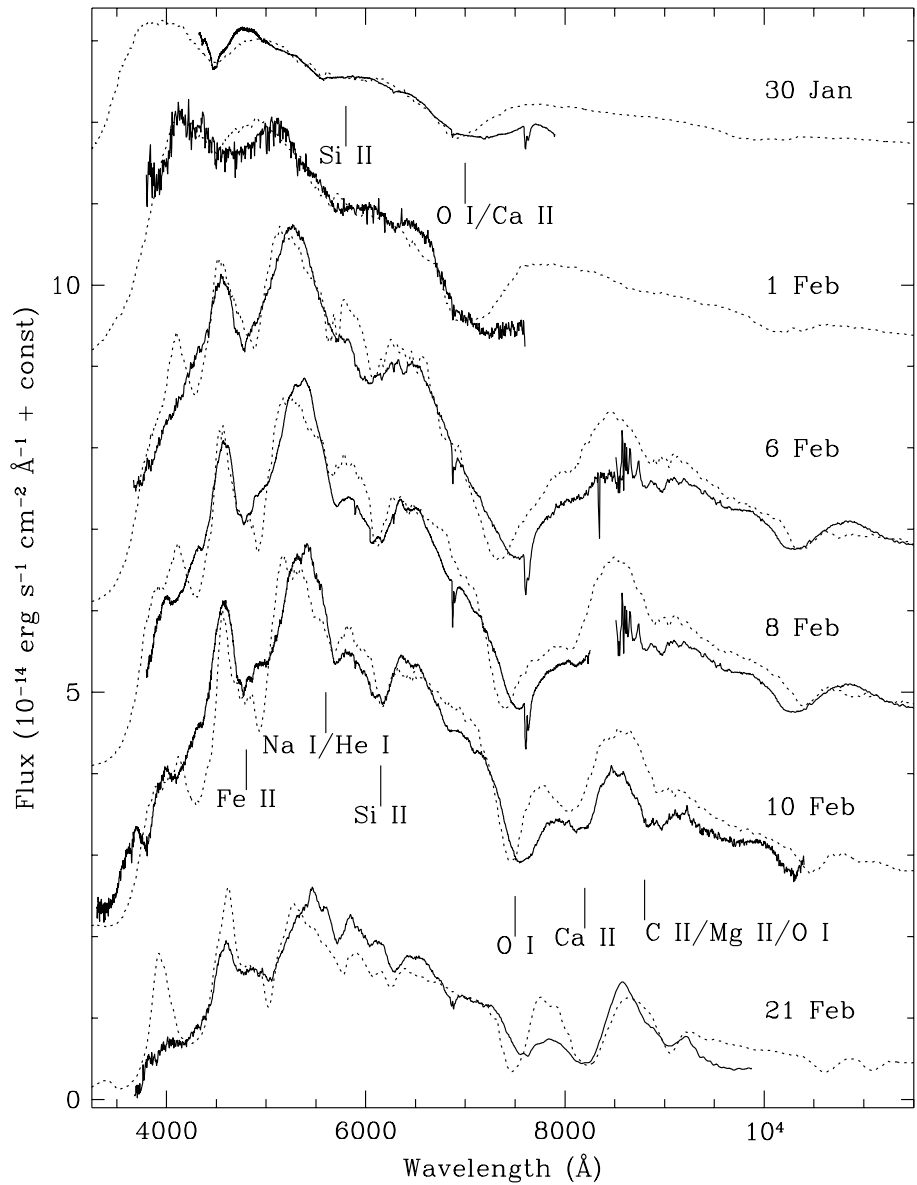}
\figcaption[.ps]{A comparison between some observed spectra of SN 2002ap 
({\em thick lines}: January 30 --- WHT (Meikle et al. 2002);
February 1 --- Gunma Obs. (Kinugasa et al. 2002); February 6 ---
Beijing Obs. (Qiu et al., in preparation); February 8 --- Subaru FOCAS
(Kawabata et al. 2002); February 7 IR --- Subaru CISCO (Motohara et
al. 2002), shown twice; February 10 --- Lick Obs. (Filippenko et al. 2002); 
February 21 --- Asiago Obs. (Turatto et al., in preparation); )
and synthetic spectra computed with model CO100/4 ({\em dashed lines}). }
\end{figure*}

\newpage
\begin{figure*}
\plotone{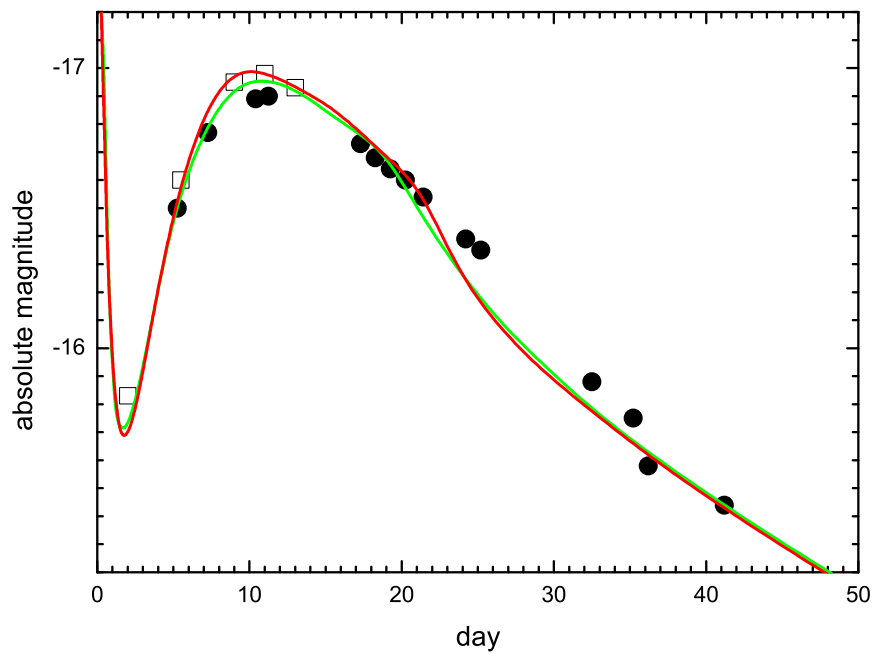}
\figcaption[.ps]{The bolometric light curve of SN 2002ap constructed using
mainly MAGNUM $UBV(RI)_cJHK$ photometry ({\em filled circles}; five points are 
based on photometry at Wise Observatory (Gal-Yam \etal 2002b), the U.K. 
Infrared Telescope (Mattila \etal 2002), MDM (Gerardy et al., in 
preparation) and Subaru Telescope (Motohara \etal 2002)).   
This is compared with two best-fit model light curves ($M_{\rm ej}=2.5$ \Msun\ 
and $E_{51}=4$: {\em red line}; $M_{\rm ej}=5$ \Msun\ and $E_{51}=8$: {\em 
green line}). 
The {\em open squares} are luminosities from the synthetic spectra. }
\end{figure*}

\end{document}